\pdfoutput=1

\documentclass[pdftex]{article}
\usepackage{icrctc07}

\title{Direct Measurements, Acceleration and Propagation of Cosmic Rays}
\shorttitle{Origin and Propagation of Cosmic Rays}

\authors{Pasquale Blasi}
\shortauthors{Blasi}
\afiliations{INAF/Osservatorio Astrofisico di Arcetri\\
Largo E. Fermi 5, 50126 Firenze (Italy)}
\email{blasi@arcetri.astro.it}

\abstract{This paper summarizes highlights of the OG1 session of the
  30th International Cosmic Ray Conference, held in Merida (Yucatan,
  Mexico). The subsessions (OG1.1, OG1.2, OG1.3, OG1.4 and OG1.5)
  summarized here were mainly devoted to direct measurements,
  acceleration and propagation of cosmic rays.
}

\begin{document}
\maketitle

\section{Prologue}

The OG1 session was splitted in 5 subsessions, devoted to direct
measurements of cosmic rays by balloons and satellites (OG1.1), cosmic
ray composition (OG1.2), cosmic ray propagation (OG1.3), cosmic ray
acceleration (OG1.4) and intrumentations and new projects (OG1.5). 
The number of papers discussed in the OG1 session gives a feeling of
the huge number of results presented during the conference: there were
138 papers presented (66 of which were oral presentations). About 37
of the oral presentations reported on observational results, while about
29 reported on theoretical or phenomenological work. It is obvious
that a meaningful summary of this work implies a selection of
highlights. An apology is due to all those that will not see
their work properly discussed here. It should also be understood that
in order to avoid doing a simple list of the results it is needed to
put things in context, and this often requires weighing the results
according to the opinions of the writer, which do not coincide
necessarily with the opinions of the Community at large. I hope that
this may be a starting point for further discussion on the topics that
I will touch upon below. 

\section{Introduction}

The origin of cosmic rays is a problem that has been haunting
scientists for almost one century now. Solving this scientific problem
means putting together numerous pieces of a complex puzzle, in
which the acceleration processes, the inner dynamics of the sources,
the propagation, the chemical composition all fit together to provide
a satisfactory and self-consistent global picture. We are not there
yet, though an increasingly larger number of pieces are finding their
place in the puzzle. The International Cosmic Ray Conference is a
privileged place to observe the puzzle taking shape.  

This rapporteur paper is organized as follows: in \S \ref{sec:bird} I
present a subjective bird's eye view of how things stand in the
field, in order to make it easier for the reader to put in the right
context the extremely wide range of results presented at the
Conference. In \S \ref{sec:measure} I summarize some important
observational results about spectra and chemical composition of cosmic
rays (including the electronic component) that have been presented. In
\S \ref{sec:accel} I illustrate some recent developments in the
understanding of the acceleration of cosmic rays. In \S 
\ref{sec:prop} I discuss the problem of relating the observed cosmic
rays to the sources through propagation in the Galaxy. The transition
from the galactic component to the extragalactic cosmic ray component
is briefly discussed in \S \ref{sec:trans} together with some issues
related to the propagation of ultra high energy cosmic rays in the
Galaxy and in the intergalactic medium. I conclude in \S
\ref{sec:concl}.  

\section{A bird's eye view on cosmic rays}
\label{sec:bird}

The best known property of cosmic rays is their all-particle
spectrum. Most experiments agree, at least qualitatively, that the
spectrum consists of four regions: 1) at low energies (below $\sim 10$
GeV) the observed spectrum is flat as it is affected by solar
modulation. 2) For energies $10\rm GeV\leq E \leq 3\times 10^{15}\rm
eV$, the spectrum can be fit with a power law with slope $\sim
2.7$. 3) At energies between $3\times 10^{15}\rm GeV$ and $\sim
10^{18}\rm eV$ 
the slope grows to $\sim 3.1$. 4) The upper boundary of region 3
is not well defined, so I will take some liberties
in defining region 4. In terms of Physics this reflects our
uncertainties in defining the region of transition to extragalactic
cosmic rays (see \S \ref{sec:trans}).

The chemical composition of cosmic rays is a crucial piece of
information and is becoming well determined at energies below the
knee, where direct measurements can still be carried out, while it is
more uncertain and model dependent at higher energies. The composition
of low energy cosmic rays provides important hints to the acceleration
processes and the propagation of cosmic rays through the interstellar
medium (ISM). Especially important in this respect are the abundances
and spectra of elements such as Boron, Beryllium and Lithium, which
are mainly produced as secondaries of primary cosmic rays. The ratio
of secondary to primary (for instance B/C) cosmic ray fluxes provides
a unique tool to characterize the diffusion properties of the
ISM. Existing measurements of this ratio as a function of energy
suggest that the diffusion coefficient scales with energy as
$D(E)\propto E^\alpha$, with $\alpha\approx 0.6$, at least
at rigidities below $\sim 10$ GV, while it is not clear whether at
higher energies the slope remains constant or there is a flattening. 
 
Elements such as Ge and Ga can potentially allow us to discriminate
between volatility based and first ionization potential based
acceleration processes, and therefore provide information about the
dominant acceleration sites. Other elements (e.g. $^{59}Co$ and
$^{59}Ni$) also provide us with precious information on the mean time
between the production of the material which is accelerated and the
actual time when it is accelerated to cosmic ray energies.

Current direct measurements are filling the gap to the knee region,
thereby providing a way to match and cross-check the indirect
measurements of spectrum and composition carried out by observing
and modelling extensive air showers at energies across and above the
knee. The data collected by the KASCADE experiment suggest that the
knee in the all-particle spectrum may be an artifact of the
superposition of sharper knees in the single components, mainly in the
light chemicals, such as H and He. The proton spectrum measured by
KASCADE has a pronounced knee at energy $\sim 10^6$ GeV. A knee in the
He component is also seen, but the fluxes become more uncertain
while moving to larger masses. It is not known as yet what could be
the explanation for the knees in these components, though it is
plausible that they may reflect inefficiency of the accelerator. In a
strictly rigidity-dependent approach, a knee in the iron spectrum
could be expected at $\sim 3\times 10^7$ GeV. This simple argument
opens the way to an equally simple but far reaching implication: the
spectrum of Galactic cosmic rays should end at energies around $\sim
10^{17}$ eV, so that cosmic rays of larger energies must be
accelerated in extragalactic sources. Unfortunately the KASCADE
results are not fully consistent with those of some other experiments,
especially the Tibet array. 
The main difficulty of all ground based experiments, including those
operating in the ultra high energy range, is the poor understanding of
cosmic ray interactions in the atmosphere, which may be very problematic
to infer basic information as the chemical composition of the primary
particles. 

Three sets of measurements will help us figure out whether the
physical picture
suggested by the KASCADE data is correct (assuming that the data
themselves find proper confirmation): 1) direct measurements of the
cosmic ray spectra and chemical composition should extend as far as
possible towards the knee and possibly across it; 2) additional data
and reliable models for the description of the shower developments
(possibly checked versus future LHC data) are crucial at energies
around and above the knee; 3) the chemical composition in the energy
region between $10^{17}$ eV and $10^{19}$ eV needs to be measured
reliably.    

Fortunately, important steps ahead in all these directions are being 
done and some important results will be discussed below.  

What about the sources? The paradigm based on supernova remnants
(SNRs) as the main sources of galactic cosmic rays remains the most 
plausible, but the smoking gun that would turn the paradigm into a
well established theory is still missing. The most important news in
this direction is represented by the recent observations of SNRs in
gamma rays by Cherenkov imaging telescopes (see \cite{funk} and
references therein for a recent review) and
the high resolution observations of the X-ray emission from the rims
of several SNRs (see \cite{jacco} and references therein for a recent
review). The former have presented us with some
evidences of TeV gamma ray emission, possibly of hadronic
origin. The latter have provided us with strong evidence of efficient
magnetic field amplification, which is in turn required in order to
reach the high energies observed in cosmic rays, and are a consequence of
efficient cosmic ray acceleration. From the theoretical point of view,
there have been several new developments in our understanding of the
mechanism of diffusive shock acceleration in SNRs but also related to
the propagation of cosmic rays both in the Galaxy and in the
intergalactic medium. I will discuss some of these issues in more 
detail below.

\section{New Measurements}
\label{sec:measure}

The direct measurements of the flux of cosmic rays and of their
chemical composition, carried out by using balloons and satellites,
has always played a crucial role in advancing our understanding of
both acceleration and propagation of cosmic rays, at least at energies   
below the knee. At higher energies the low fluxes due to the steeply
falling spectrum make it necessary to use ground arrays which observe
cosmic ray induced air showers. The two techniques are clearly
complementary and one of the biggest problems has always been to
cross-calibrate the two. For the first time this 
goal seems to be at least in sight, in that the direct measurements are
extending to $\sim 10-100$ TeV, therefore approaching the knee
region. Numerous new results on spectra of different chemical elements
have been presented, from balloon flights (CREAM, ATIC, Tracer, TIGER,
Bess-Polar, PPB-BETS), from satellites (preliminary results from
PAMELA) and even some shower experiments (Tibet Array and HESS).  
The overlap between all these techniques in the energy region around
the knee appears to be of crucial importance, especially to unveil the 
origin of the knee in the cosmic ray spectrum. 
\vskip .5cm
{\it Balloons and Satellites}

\begin{figure*}
\noindent
\begin{center}
\includegraphics [width=.45\textwidth]{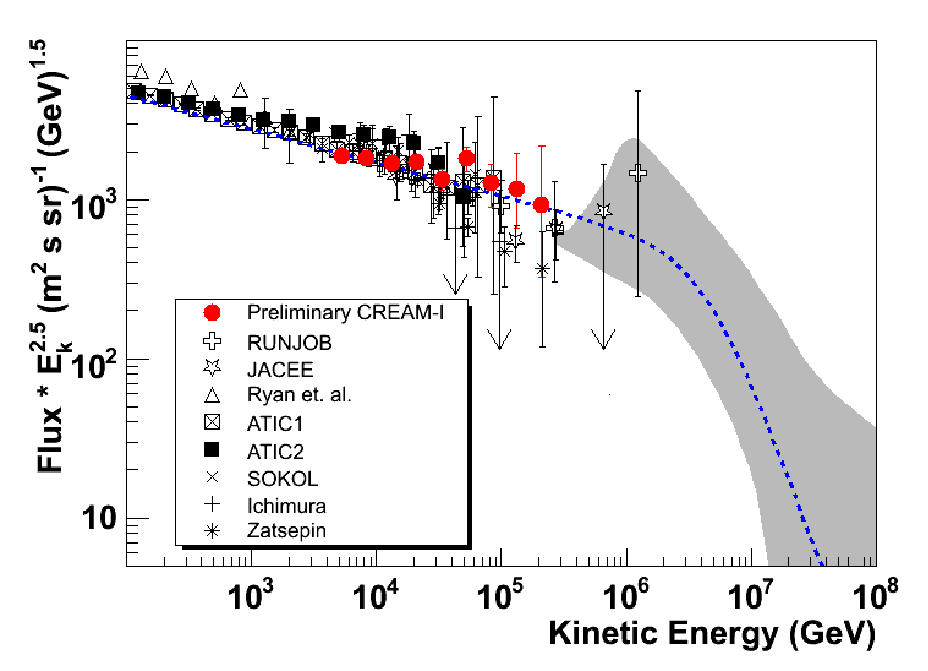}
\includegraphics [width=.47\textwidth]{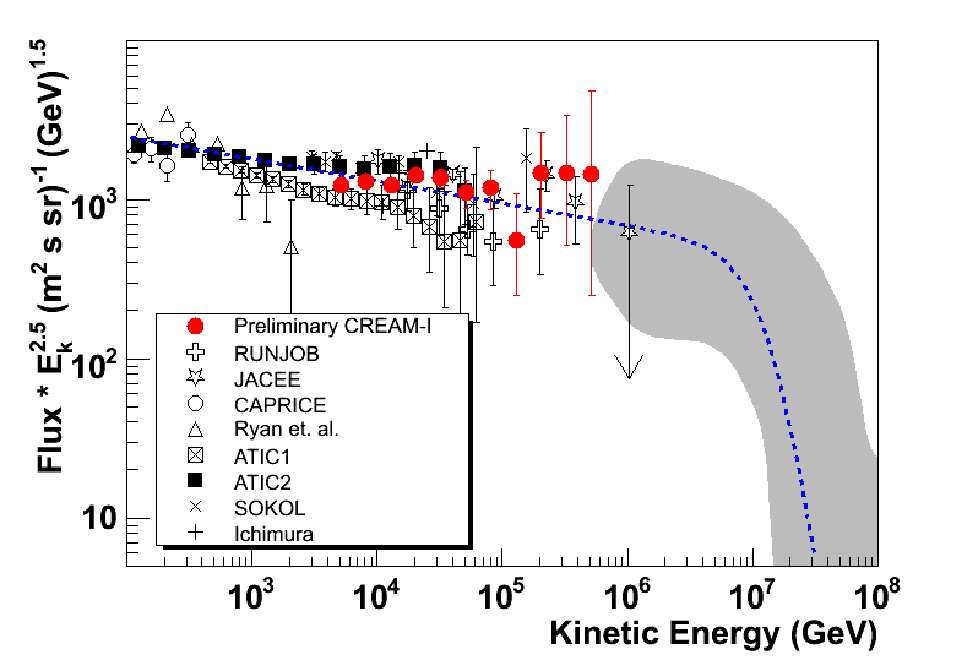}
\end{center}
\caption{Spectra of protons (left panel) and helium nuclei (right
  panel) as measured by CREAM-I (figure from \cite{seo0677}) compared
  with the results of other experiments.}
\label{fig:cream}
\end{figure*}

The CREAM Collaboration has presented impressive results collected
during a total of 70 days of flight of their balloon experiment. The
data were collected during the record breaking CREAM-I flight (42 days
between 12/16/05 and 01/27/05) and a second flight (CREAM-II) lasted
28 days (between 12/16/05 and 01/13/06). Convincing evidence was
presented of the excellent charge resolution of the experiment
($\Delta q\sim 0.2$ electron charge), precious tool to reconstruct the
spectra of different chemicals (see \cite{parketal,coutuetal} for a
technical discussion). 

In Fig. \ref{fig:cream} I reproduce the plots as shown by
\cite{seo0677} (see also \cite{yoon0778}) illustrating the spectra of
protons and helium nuclei as measured by CREAM-I (red circles). The CREAM
spectra are compared with those of other experiments as listed in the
figure with different symbols. The spectrum of He nuclei as measured
by CREAM is compatible with that measured by ATIC-2, but appears to be
somewhat flatter than that measured by ATIC-1. 

The spectra of Carbon and Oxygen
measured by CREAM-II were presented by \cite{zei0301} and
\cite{ahn1055}. The results on the C/O ratio confirm the primary
nature of both nuclear species \cite{park0593}. An overview of the
CREAM results and future developments (a third and fourth flight) were
presented by \cite{seo0677}. The spectra of the chemicals presented by
the CREAM collaboration do not show appreciable differences in the
slope, with the possible exception of the helium spectrum that might
be slightly harder than the proton spectrum. In fact this trend is
probably present even in the ATIC-2 spectrum. In Fig. \ref{fig:cream}
the shaded area is supposed to show the range of fluxes as measured by 
ground arrays. The width of the shaded region provides an estimate of
the uncertainties due to either the interaction models adopted for the
analysis or systematics in the different experiments.
An important point that
this figure shows is that finally direct and indirect measurements 
are starting to overlap in the knee region. This will be of great
importance to understand the reason for the appearance of the knee in
the all-particle cosmic ray spectrum. 

The ATIC balloon experiment has presented impressive results on the B/C
ratio and mass composition of cosmic rays below the knee. In fact the
ratios N/O and C/O were also measured. The B/C ratio, as I discuss in
\S \ref{sec:prop}, is a crucial indicator of the mode of cosmic ray
propagation in the interstellar medium. Low energy measurements
($E<30$ GeV) carried out by HEAO-3 \cite{heao3} show that the B/C
ratio scales with rigidity as $\sim R^{-0.6}$, usually interpreted as
a proof of the primary (secondary) origin of Carbon (Boron) nuclei. 
\begin{figure}
\noindent
\includegraphics [width=.5\textwidth]{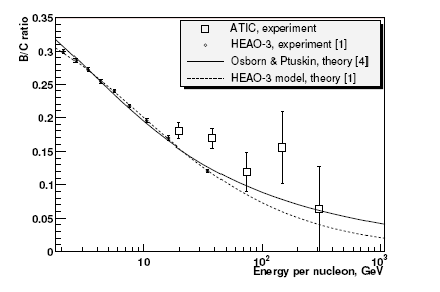}
\caption{B/C ratio as a function of energy per nucleon as measured by
  ATIC-2 (squares) \cite{panov0018} and compared with the results of
  HEAO-3 and the predictions of two theoretical models (lines).}
\label{fig:atic_bc}
\end{figure}

The ratio as measured by ATIC-2 is reported in Fig. \ref{fig:atic_bc}
\cite{panov0018}
together with the previous results of HEAO-3 and the predictions of
two theoretical models (lines) (see \cite{panov0018} for some
discussion and references). Despite the apparent flattening of the
measured B/C ratio in the ATIC data, the error bars are still too
large to infer solid conclusions on this issue.  

\begin{figure*}
\noindent
\includegraphics [width=.5\textwidth]{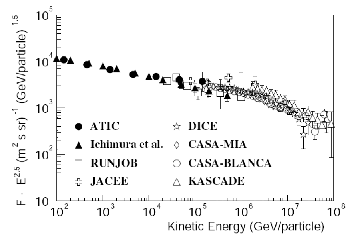}
\includegraphics [width=.5\textwidth]{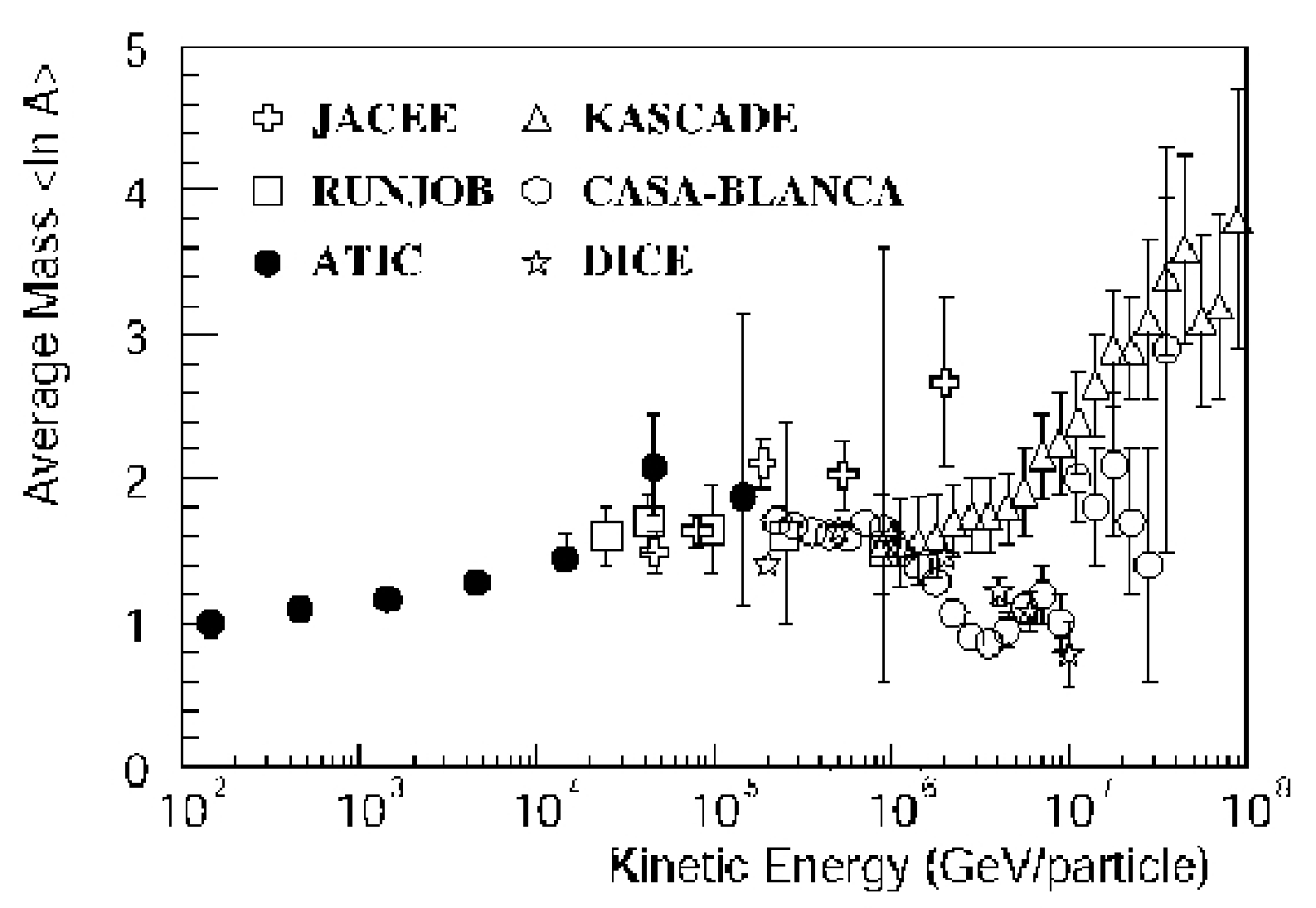}
\caption{All particle spectrum (left panel) and mean logarithmic mass
  number (right panel) as measured by ATIC \cite{ahn1173}.}
\label{fig:atic}
\end{figure*}

The all-particle spectrum and the average mass number as a function of   
energy as measured by ATIC-1 are shown in Fig. \ref{fig:atic}
\cite{ahn1173} compared with the results of other experiments, as
indicated in the figure. There is a substantial agreement among all
the data sets shown. At least in part the small offsets may be
explained in terms of systemaic errors in the energy determination,
amplified by the multiplication by $E^{2.5}$. 

The mean logarithmic mass number found by ATIC in the energy range
$10^2-10^5$GeV shows a general trend to a heavier composition and
appears to match well with the results of RUNJOB, CASA-BLANCA, DICE
and KASCADE in the knee region. At energies above the knee (not
reached by ATIC) the other experiments show quite different
trends. This prevents us from reaching a satisfactory
explanation of the origin of the knee and of the transition from
galactic to extragalactic cosmic rays (see \S \ref{sec:trans}).

\begin{figure}
\noindent
\begin{center}
\includegraphics [width=.5\textwidth]{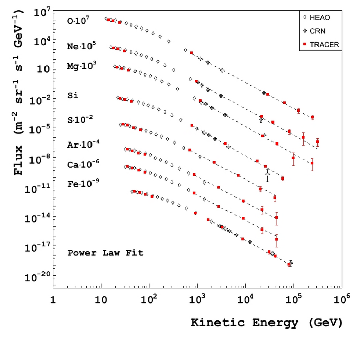}
\end{center}
\caption{Differential spectra of chemical elements from Oxygen to Iron
  as measured by Tracer \cite{boyle1192} and compared with previous
  results from CRN and HEAO-3.}
\label{fig:tracer}
\end{figure}

Very accurate measurements of the flux of different chemical elements
below the knee were presented by the TRACER Collaboration
\cite{boyle1192} and are shown in Fig. \ref{fig:tracer} (filled dots)
for nuclei between Oxygen and Iron.  
The fluxes are multiplied by different normalization factors to make
the plot clearer, and they are compared with the results of HEAO-3 and
CRN. No appreciable difference between the slopes of the spectra of
these nuclei was detected, all slopes being around $\sim 2.7$. 

\begin{figure}
\noindent
\includegraphics [width=.5\textwidth]{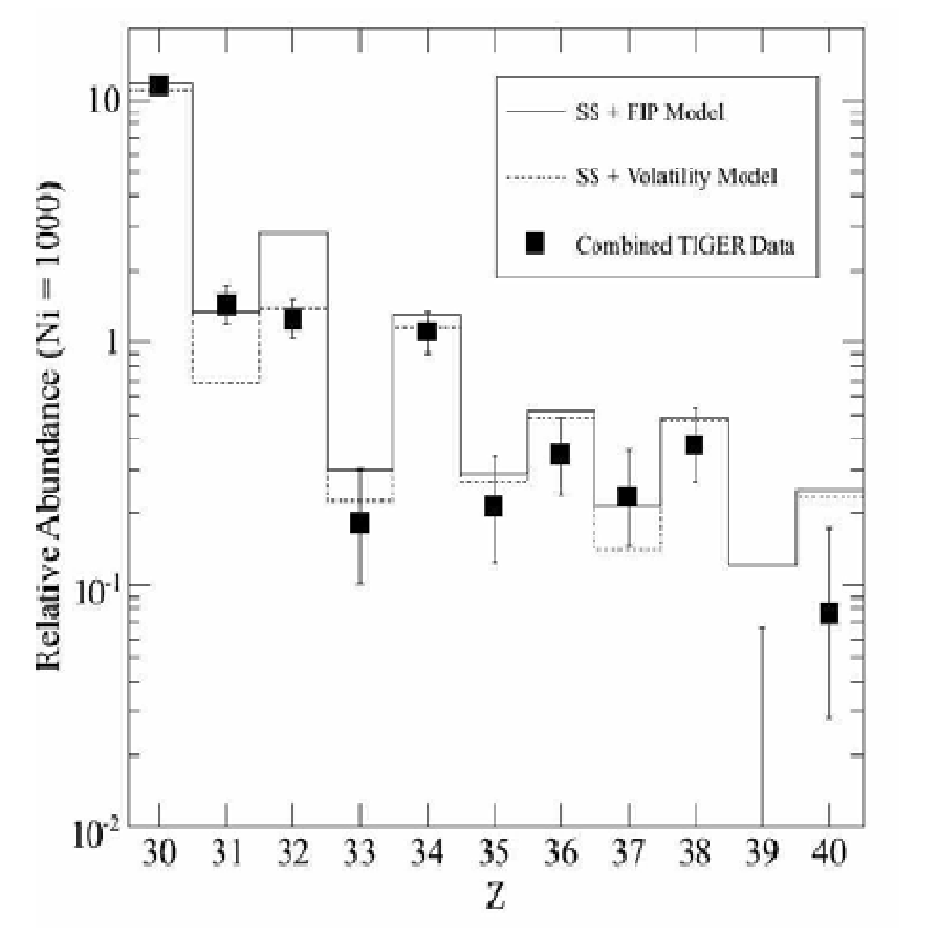}
\caption{Relative abundances of heavy elements as observed by the
  TIGER experiment \cite{rauch0187}.}
\label{fig:tiger}
\end{figure}

The TRACER Collaboration also presented the B/C ratio in the energy
region around $\sim 100$ GeV/amu, being fully consistent with other
measurements. The predicted error bars on the B/C ratio with a 30 days
balloon flight of TRACER should allow to finally pin down the slope of
the ratio as a function of energy up to $\sim 1$ TeV/amu. 
While indicators such as the B/C ratio provide us with important
information on the propagation of cosmic rays in the Galaxy, the
abundances of super-heavy elements ($Z>30$) tell us about the
acceleration regions, though their flux is more than $\sim 1000$ times
smaller than the Iron flux. The flux of these elements has been
measured during two balloon flights by the TIGER Collaboration in 2001
and 2003 (a total of about 50 days of observations). The fluxes in the
energy region between $0.3$ and $10$ GeV/nucleon are reported in
Fig. \ref{fig:tiger} (from \cite{rauch0187}). The lines represent the
predictions of two 
models for acceleration, one based on first ionization potential (FIP)
and the second on volatility, but both with a standard solar system
composition of the ambient medium. The abundances of $^{31} Ga$ and
$^{32} Ge$ appear to be inconsistent (if taken at the same time) with
both theoretical models. One should however recall that a standard
solar system composition near the accelerator is all but granted. 

Important information on the origin of cosmic rays and their
propagation can also be indirectly gathered by observing electrons.    
Their propagation in the Galaxy is affected by diffusion, as for ions,
and energy losses (mainly synchrotron and inverse Compton scattering
losses). At energies larger than 20-100 GeV the loss time due to ICS
and synchrotron becomes shorter than the diffusive escape time from the
Galaxy, so that the spectrum steepens by one power compared with the
source spectrum. At energies larger than $\sim 1$ TeV the flux of
electrons at 
Earth is expected (and observed) to decline, presumably as a result of
the discreteness in the spatial distribution of the sources. In this
sense several groups have been engaged in a search for a possible
signal of the closest sources of electrons around the solar
system, the most promising sources being Vela, the cygnus loop and
Monogem. The expected signal would consist of a bump in the diffuse flux of
electrons at energies in excess of $\sim 1$ TeV. The proximity of the
source would also result in a small anisotropy in the arrival
directions. The spectrum and anisotropies have been presented by
PPB-BETS.
\begin{figure}
\noindent
\includegraphics [width=.5\textwidth]{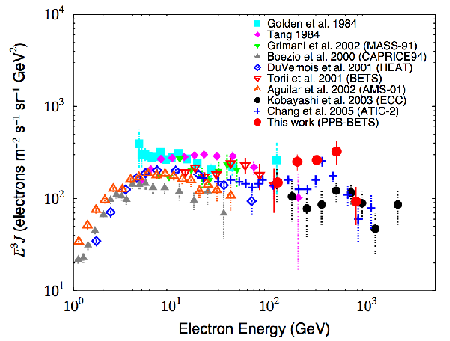}
\caption{Spectrum of electrons (multiplied by $E^3$) as observed by
  PPB-BETS \cite{yoshida0892} compared with the results of previous
  measurements.} 
\label{fig:PPBspec}
\end{figure}

To date, the spectrum \cite{yoshida0892} (Fig. \ref{fig:PPBspec}) does not
show evidence for the possible appearance of a nearby electron
source. Even the anisotropy is claimed to be fully consistent with an
isotropic distribution of arrival directions \cite{yoshida0892}.  
\begin{figure}
\noindent
\includegraphics [width=.5\textwidth]{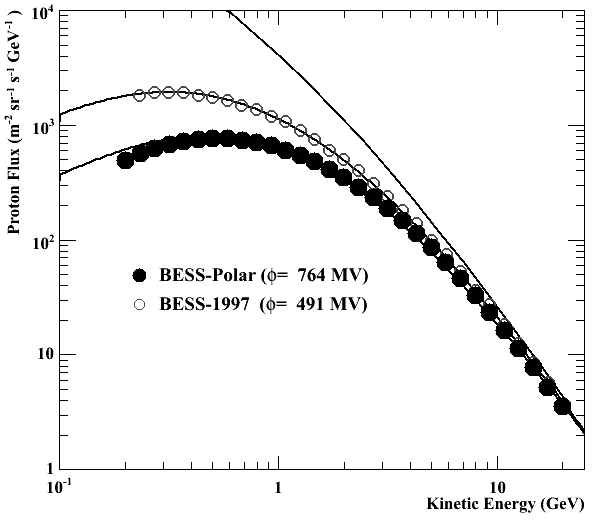}
\caption{Spectrum of protons as observed by BESS-1997 (empty circles)
  and BESS-Polar (filled circles) \cite{hams1119}. The difference
  between the two spectra is due to the different levels of solar
  modulation at the time of the measurements.}
\label{fig:BESSspec}
\end{figure}

The BESS-Polar experiment has presented several interesting results,
ranging from low energy cosmic ray spectra to spectra of antiprotons
and limits to the flux of anti-helium nuclei. The flux of cosmic ray
protons in the energy region below $\sim 10-30$ GeV is heavily
affected by solar modulation. In Fig. \ref{fig:BESSspec} I report the
results presented by \cite{hams1119}: the thick dots are the recent
results of BESS-Polar, compared with those of BESS-1997 (open
circles), at the time of the solar minimum. It is evident the effect
of the enhanced solar modulation while the sun gradually approaches
the next solar minimum. The upper solid line represents the spectrum
of protons in the interstellar medium as inferred from BESS-98
measurements. 

\begin{figure*}
\noindent
\begin{center}
\includegraphics [width=.85\textwidth]{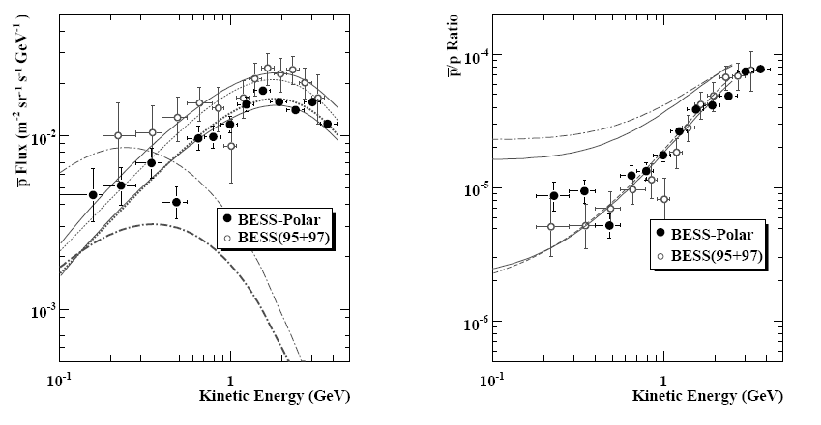}
\end{center}
\caption{Left Panel: Antiproton spectrum measured in BESS-Polar
compared with BESS(95+97) results. The lines illustrate predictions of
different theoretical scenarios. The lines that pass through the data
points are the predictions of standard propagation calculations
(GALPROP and leaky box models). Right Panel: ${\bar p}/p$ ratio
measured by BESS(95+97) and BESS-Polar I (2004) compared with drift
model calculations at various Solar magnetic field tilt angles.}
\label{fig:BESSantip}
\end{figure*}

\begin{figure}
\noindent
\includegraphics [width=.5\textwidth]{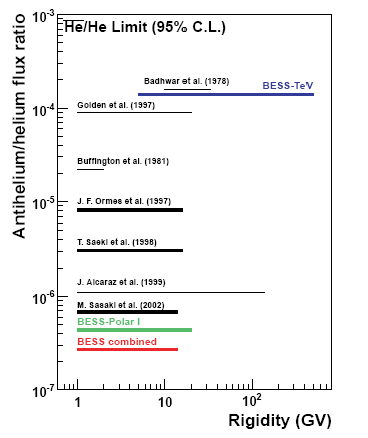}
\caption{Limits on the flux of anti-Helium obtained by BESS-Polar
  \cite{hams1119}.}
\label{fig:antiHe}
\end{figure}

The propagation of cosmic ray nuclei in the interstellar medium also
results in the production of many secondary products (nuclei,
antiprotons, electron-positron pairs, gamma rays, neutrinos). Accurate
measurements of these secondaries, especially antiprotons and
positrons is instrumental to the search for faint signals, such as
those coming from the annihilation of non-baryonic dark matter. The
flux of anti-protons and the ratio of anti-protons to protons in the
energy region below few GeV as measured by BESS-Polar is reported in
Fig. \ref{fig:BESSantip}. In the left panel, the antiproton spectrum
is compared with some theoretical predictions of the same quantities
for some models (solid line: propagation with GALPROP; dotted line:
standard leaky box with solar modulation; dash-dotted line: black hole
evaporation). In the right panel, the ratio antiproton/proton flux is
compared with predictions of a drift model with different tilt angles
of the solar magnetic field. The thick dots and empty circles refer
again to BESS-Polar and BESS-1997 respectively. In the right panel the
effect of a different solar modulation in the two periods is evident
again. As expected, this effect almost completely disappears in the
ratio of the antiproton/proton flux. The fluxes of antiprotons
observed by BESS-1997 and BESS-Polar are both fully consistent with
the one expected on the basis of a propagation model with solar
modulation, thereby implying significant upper limits to the flux of 
antiprotons from, for instance, neutralino annihilation in the
Galactic dark matter halo. 
BESS has also focused on the detection of anti-nuclei. The quest for
why our universe is (almost) completely made of matter instead of a
mixture of matter and anti-matter is a fundamental one, as Nature is
expected to have produced a (almost) symmetrical universe. Though it
is likely that the small excess of matter over anti-matter is what now
makes most of the present universe, the search for islands of
antimatter or traces of anti-nuclei has never quite finished. The
limit imposed by BESS-Polar and the limit that will possibly be
reached by the next BESS flight are shown in Fig. \ref{fig:antiHe},
together with the results of previous measurements. 

\begin{figure*}
\noindent
\begin{center}
\includegraphics [width=.65\textwidth]{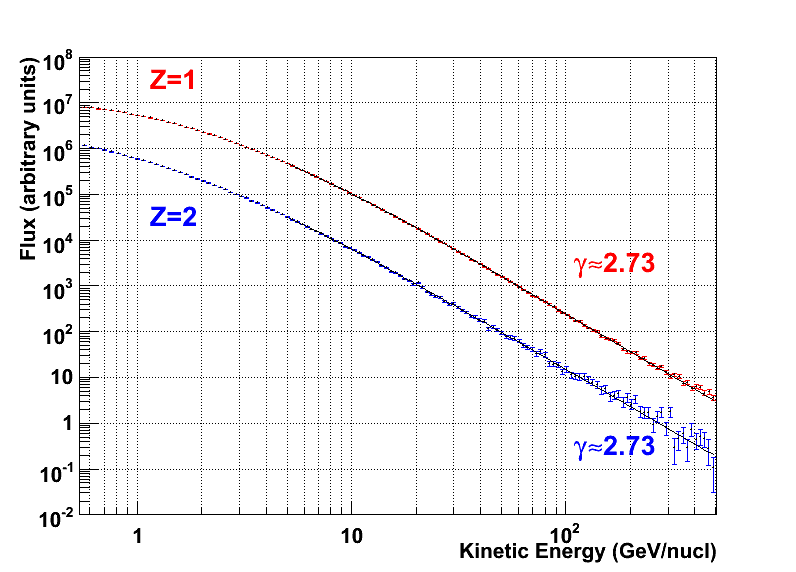}
\end{center}
\caption{Preliminary spectra of protons and helium from PAMELA.}
\label{fig:PAMELA}
\end{figure*}

The measurement of the cosmic ray spectrum and chemical composition up
to energies of $\sim 1$ TeV, the search for dark matter, anti-matter
and exotic particles, and finally bits of solar and magnetospheric
physics are among the goals of the PAMELA satellite, which has been
succesfully launched on june 15, 2006. The instruments onboard are
collecting data, but only preliminary results were presented at the
conference. A review of PAMELA preliminary results and expectations
has been presented in \cite{picozza0289}.
Particularly interesting, and suggestive of the quality of
the results that we should expect, is the spectrum of protons and
helium, reported in Fig. \ref{fig:PAMELA}. 
This preliminary analysis does not show evidence for any difference in
the slopes of the spectra of H and He in the energy region below $\sim
500$ GeV. In both cases the best fit power law has a slope of $2.73$.

Several simulated results were also presented by the PAMELA
Collaboration, mainly aimed at showing the potential for discovery of
indirect signals of dark matter annihilation (in the form of anomalous
features in the antiproton and positron spectra), and the limits
achievable in the search for anti-matter. The limit on the ratio
${\bar He}/He$ is expected to improve by about 1 order of magnitude in
about three years of PAMELA operation with respect to that achieved by
BESS-Polar (see Fig. \ref{fig:antiHe}). 

PAMELA is expected to play a crucial role in the identification of
spectral features in the positron and antiproton spectra as they can
possibly result from dark matter annihilation, provided the mass of
the dark matter candidate particle is in the right energy range (below
$500-1000$ GeV). 

\vskip .5cm
{\it Ground Experiments}

Since most results from ground experiments were presented in other
sessions of the ICRC, the talks on this subject in the OG1 session
were sparse. 

\begin{figure}
\noindent
\includegraphics [width=.5\textwidth]{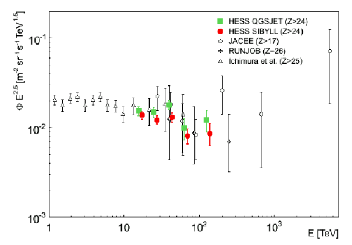}
\caption{Spectrum of iron nuclei (in fact nuclei with $Z>24$ obtained
  by HESS using the detection of DC light. The HESS results (with
  SYBILL and QGSJET as models for interactions) are compared with the
  results of other measurements.}
\label{fig:hess}
\end{figure}

The HESS Collaboration has presented the results of a very interesting
detection of the direct Cherenkov (DC) light from cosmic rays
impacting the atmosphere. This technique, proposed as a possible tool
to measure the chemical composition of cosmic rays above 10 TeV by
\cite{kieda}, has
been used by HESS to measure the flux of heavy nuclei (Fe-like). The
direct Cherenkov light is produced by the primary nucleus while
entering the atmosphere and before the first interaction that
generates the shower. High up in the atmosphere the density of air is
lower and therefore the Cherenkov cone is more collimated with respect
to the Cherenkov cone of the light generated by lower energy particles
in the shower. As a result the total Cherenkov emission from a shower
is expected to have a broad region with a {\it hot} pixel
corresponding to the DC light. This emission has in fact been
successfully detected by HESS. Since the intensity of the Cherenkov
signal scales with the square of the charge of the parent nucleus, it
is best to seach for the signal from Fe nuclei provided the energy
does not exceed a few hundred TeV, in order to avoid that the shower
emission overshines the DC light. The spectrum of Iron nuclei as
measured by HESS is reported in Fig. \ref{fig:hess} \cite{hess0284} where
it is compared with the results of other experiments. The two sets of
data points refer to QGSJET and SIBYLL as interaction models.

The search for the DC light is also being pursued by VERITAS
\cite{OG0732}, while dedicated instruments are being planned
(e.g. TrICE \cite{Trice}). 

The HESS Collaboration has also presented the preliminary results of
their measurement of the electron spectrum \cite{OG0446}, consistent
with measurements carried out by other experiments within the
systematic uncertainties. 

A relevant contribution to the understanding of the nature of the knee
in the all-particle spectrum has been provided by Tibet Array. The
Collaboration has presented the most recent measurements of the proton
and helium spectra in the knee region, and the fraction of heavy
nuclei as a function of energy, in the range $10^6\leq E\leq 10^7$
GeV. 

The measurements of the chemical abundances in this energy region are
still controversial. The Tibet data on the proton spectrum appear in
rough agreement with those of RUNJOB and JACEE (both with QGSJET and
SIBYLL as interaction models) but in rather apparent contradiction
with the KASCADE data, especially when the KASCADE data are analyzed
using SIBYLL. The latter show a steeper spectrum of protons and a
somewhat higher normalization of the flux. The discrepancies are even
more evident in the case of the helium spectrum. The Tibet results
on helium roughly match with those of RUNJOB, but not with data from
JACEE and KASCADE. 
The absolute flux of helium as measured by KASCADE is almost
one order of magnitude higher than that measured by Tibet Array. The
situation in the knee region continues to be confused and this
prevents the possibility of unveiling the origin of the knee.   

\section{Acceleration of Cosmic Rays}
\label{sec:accel}

The main mechanism for the acceleration of cosmic rays (not only
galactic) remains diffusive particle acceleration at shocks. Many
contributions presented at the conference have focused on the
determination of the efficiency of particle acceleration, the
dynamical effect of the accelerated particles and the magnetic field
amplification generated by the accelerated particles through streaming 
instability or due to turbulent amplification. 

I will start by discussing some issues related to the acceleration of
Galactic cosmic rays in SNRs, though some of the conclusions will be
of wider applicability. 

One of the most recent advancements in the theory of particle
acceleration at non-relativistic shocks, especially for SNR shocks,
has been the calculation of the dynamical reaction of the accelerated
particles onto the background plasma. The effect has been discussed
already in the 80's in the context of two-fluid models (see
\cite{maldru} for a review),
and later addressed numerically by direct solution of the time
dependent equation of diffusive tranport coupled with the equations of 
mass, momentum and energy conservation of the background plasma (see
for instance \cite{bell87}). In the late 90's a stationary
semi-analytical solution of this set of equations was found in the
form of an integral equation \cite{malkov} for a spatially constant
diffusion coefficient. A general semi-analytical method was
recently proposed by \cite{amato1}, valid for any choice of the
diffusion coefficient. 

The structure of the shock is changed by the accelerated particles due
to the pressure they exert on the background plasma, though in a
collisionless manner. For ordinary diffusion coefficients, which are
increasing functions of particle momentum, higher energy particles can
travel further away than low energy particles: a fluid element
approaching the shock surface therefore {\it feels} an increasing
pressure due to accelerated particles and as a consequence it slows
down. This leads to the formation of a {\it precursor} upstream of the
shock (which is now called {\it subshock}) in which the fluid
velocity, as seen in the (sub)shock reference frame decreases while
approaching the shock. The effective compression factor {\it felt} by
particles crossing the shock and precursor is now a function of
momentum, being bigger at large momenta and smaller at low
momenta. The immediate consequence is that the spectrum of particles
accelerated at shocks is not a power law, as expected in test-particle
theory, but rather a concave function of momentum, harder at high
momenta and softer at low momenta. This nonlinear system is found to
self-regulate itself and to be able to reach relatively high
efficiency of particle acceleration. 

This complex nonlinear system has however additional interesting
aspects, concerning the maximum momentum that the particles can be
accelerated to. As was recognized first in \cite{lc83a,lc83b}, the maximum
momentum achieveable in SNRs is exceedingly low compared with the knee
energy unless the diffusion coefficient is Bohm-like and magnetic
field amplification takes place. Even in this case the maximum energy
falls short of the knee energy by a factor $\sim 100$, unless the
amplification turns strongly nonlinear, namely $\delta B/B\sim 100$.  
An investigation of the process of particle acceleration at cosmic ray
modified shocks with self-generation of strong turbulence was
discussed in \cite{amato2,damiano} and presented at this Conference by
\cite{blasi0341}. 

The most important clue in this field has however come in the last
few years from high resolution X-ray observations of the rims of
several SNRs. The thickness of the rims is related to the loss length
of high energy electrons radiating X-rays by synchrotron emission. The
observed brightness profiles lead to estimates of the downstream
magnetic field of the order of a few hundreds $\mu G$
\cite{berevolk}. These estimates were confirmed and presented here by
\cite{berez0597,berez0614} together with their implications for
multifrequency observations of specific SNRs. 

\begin{figure*}
\noindent
\begin{center}
\includegraphics [width=.9\textwidth]{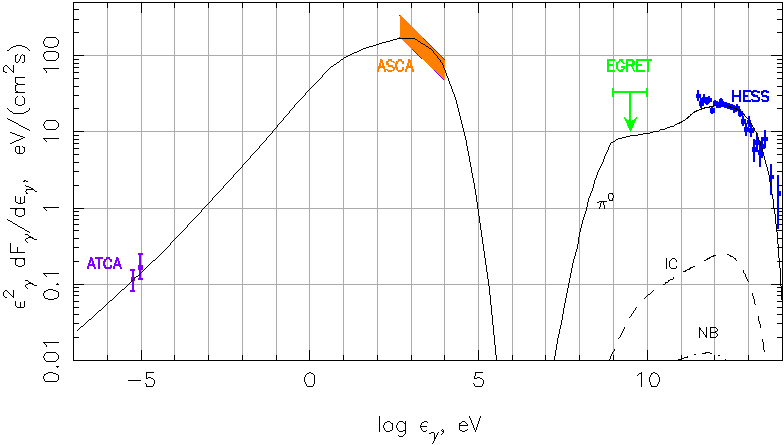}
\end{center}
\caption{Multifrequency spectrum of the SNR RXJ1713.7-3946 and the
  best fit of the calculations of \cite{berez0614}.}
\label{fig:snr1713}
\end{figure*}

These calculations are based on a numerical time-dependent solution of
the problem of particle acceleration at modified shocks. The strength
of the magnetic field is not self-consistently calculated but rather
fit to the observations at a given time (the observation time) and not
evolved in time. The calculations of \cite{berez0614} show that the
multifrequency data for supernova RXJ1713.7-3946 are explained rather
well from radio to X-rays and gamma rays (Fig. \ref{fig:snr1713} from
\cite{berez0614}). In
particular the gamma ray data from HESS are interpreted as the result
of pion production and decay. The magnetic field required for the fit
is $126 \mu G$, again consistent with the general trend of strong
magnetic field amplification observed in other remnants. 
The strong magnetic field causes the electron density to be low,
thereby reducing the flux of ICS gamma rays. Similar consequences can
be inferred for the RX J0852.0-4622 (Vela Jr.) \cite{berez0597}. It is
also worth noticing, in Fig. \ref{fig:snr1713} (and others that have
been presented for other SNRs) that both the predicted radio-X and
gamma ray spectra show a clear indication of a curvature
(concavity). This characteristic, as discussed above, is one of the
most distinctive consequences of particle acceleration at modified
shocks. Signatures like this should be carefully looked for in future,
better data at all wavelengths. In particular, future gamma ray
observations with GLAST might allow us to confirm or reject the
hadronic origin of the gamma ray emission.

Two models have been discussed for the amplified magnetic field
observed in several SNRs. Quantitatively they potentially lead to the
same level of magnetization, but the principle is different in
the two cases. 

The traditional explanation is based on streaming instability induced
by the accelerated particles. The instability leads to resonant growth
of Alfven modes \cite{bell78} and typically $\delta B/B\sim
20-30$ \cite{amato2}. By using the results of quasi-linear theory for
the growth rates, \cite{blasi0341} showed how to determine the
diffusion coefficient and implement it in a self-consistent solution
of the problem of acceleration at modified shocks. At the beginning of
the Sedov phase this type of instability leads to reach approximately
proton energies of $\sim 10^6$ GeV, while
at later stages of the SNR evolution, the maximum momentum decreases
in a way that depends on how the magnetic field amplification drops. 

\begin{figure}
\noindent
\includegraphics [width=.5\textwidth]{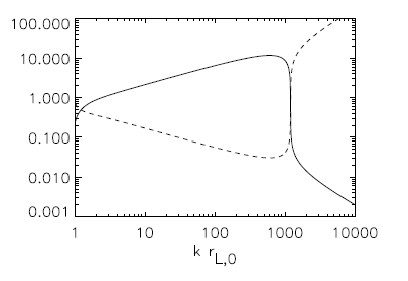}
\caption{Real and Imaginary part of the frequency of the waves excited
by the streaming of cosmic rays at a shock with velocity $10^9\rm
cm~s^{-1}$ and efficiency of particle acceleration $\eta=0.1$. For
these parameters the peak of the growth rate occurs at $k r_L\gg 1$,
showing the non resonant nature of the instability \cite{blasi0342}.}
\label{fig:disp}
\end{figure}

In \cite{bell2004} it was proposed that non resonant modes may grow
faster than resonant modes in some circumstances. The analysis of Bell
was based on a MHD treatment of the background plasma and the non
resonant modes are often considered as strictly related to this
assumption. The presentation of \cite{blasi0342} has demonstrated
that the same dispersion relation is obtained in the context of a
purely kinetic approach, if particle acceleration is efficient, as it
is expected at cosmic ray modified shocks. The imaginary and real
parts of the frequency of the propagating waves are plotted in
Fig. \ref{fig:disp} in units of 
$V_s^2/c r_{L,0}$ as a function of the wavenumber $k$ in units of
$1/r_{L,0}$. Here $r_{L,0}$ is the Larmor radius of the particles in
the background magnetic field and $V_s$ is the shock velocity, taken
here as $10^9\rm cm~s^{-1}$. The efficiency in acceleration was
assumed to be $\eta\equiv U_{CR}/\rho V_s^2=0.1$.
The peak of the growth rate if not at $k r_{L,0}\sim 1$ which is again
a confirmation that the faster growing modes are non resonant. However
the same authors of  \cite{blasi0342} also showed that the peak moves
towards $k r_{L,0}\sim 1$ when either the shock velocity decreases or
$\eta$ decreases. This suggests that at later epoques during the SNR 
evolution the resonant mode may become dominant.

The search for the non resonant modes found by \cite{bell2004} was
also carried out by using Particle-in-Cell (PIC) simulations. The
results of such studies were presented by \cite{niemiec1047}: no evidence
was found of the fastly growing {\it Bell mode} in their simulations. 
It remains to
be seen whether this result is due to the anomalous values of the
parameters adopted by the authors (for instance the large density of
non thermal particles needed to satisfy the condition of efficient
acceleration and the artificial value of the ratio $m_e/m_p$), or if
actually shows that the instability may be suppressed due to some
physical mechanism yet to be identified. 

It is important to stress that streaming instability generates the
magnetic field amplification upstream of the shock front. The
turbulent field is then advected downstream and the perpendicular
components are compressed at the shock surface. 

\begin{figure}
\noindent
\includegraphics [width=.5\textwidth]{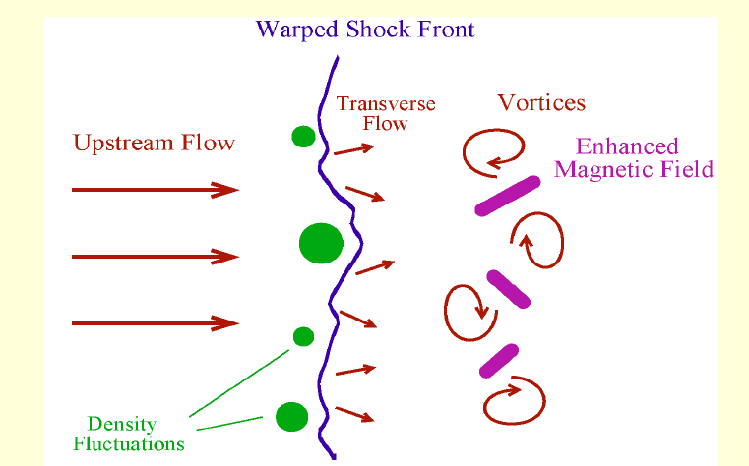}
\caption{Schematic view of the turbulent magnetic field amplification
  scenario proposed by \cite{joki0078}.}
\label{fig:joki}
\end{figure}

A second model for the generation of strong field was discussed by
\cite{joki0078}: the model is based on the (realistic) assumption that
density perturbations $\delta \rho/\rho\sim 1$ are present
upstream (see Fig. \ref{fig:joki} for a schematic view). The density
perturbations induce corrugations in the shock structure which in turn
produce turbulent eddies that twist the field lines frozen in the
plasma leading to magnetic field amplification. The numerical
simulations that the authors illustrated show that $\delta B/B\sim
M_A$, where $M_A$ is the Alfven Mach number. In conditions which are 
typical of SNRs one may easily expect $\delta B/B\sim 100-1000$
downstream of the shock. 
The effects of such amplified magnetic field on the acceleration of
particles need some further discussion: if the field in the upstream
region is mainly parallel to the shock normal, and there is no
streaming instability (as assumed in \cite{joki0078}) then the
acceleration process is inefficient, being weakly affected by the
downstream strong fields alone. In fact the acceleration time is the
sum of the diffusion times upstream and downstream. If the field is
only amplified downstream, the acceleration time is dominated by the 
upstream region, and it remains too long to lead to efficient
acceleration. 

The problem can be avoided by assuming that the shock is
perpendicular. In this case the
upstream acceleration time is substantially shorter because the
perpendicular diffusion coefficient is smaller than the parallel one
\cite{perp}, 
thereby mimicking amplification of the magnetic field from the point
of view of acceleration. 
For perpendicular shocks the amplification by streaming instability is
expected to be suppressed, therefore the two mechanisms could be
considered as complementary and might take place together though with
different efficiencies as related to the relative orientation of the
shock normal and the ambient magnetic field. 

Efficient particle acceleration at SNR shocks modified by the
dynamical reaction of cosmic rays, together with the amplification of
the magnetic field provide a consistent picture of the origin of
cosmic rays: protons in the SNR environment can be accelerated to
$\sim 10^6$ GeV, while elements with higher charge ($Z$) may be
accelerated to energies which are $Z$ times larger. The maximum
energies are reached at the beginning of the Sedov phase while at
later stages the maximum energy decreases, mainly because the magnetic
field amplification is less efficient (at least for amplification due
to streaming instability). The spectrum of cosmic rays observed at the
Earth is the superposition of the cosmic rays trapped behind the shock
and released after adiabatic decompression and the particles which may
leave the system from upstream due to lack of confinement in the
accelerator. 

The calculations presented by \cite{berez0111} represent the first
attempt to convolve the results from particle acceleration with given
diffusion properties of the interstellar medium to determine the
all-particle spectrum of Galactic cosmic rays. Good agreement with
data is obtained after summing over all chemical elements, though
there are numerous parameters that can be tuned to obtain the fit. The
most important point is however that the end of the Galactic spectrum
is expected to be at energies around a few $10^{17}$ eV. I will
discuss this point in more detail in \S \ref{sec:trans}. 

Our understanding of cosmic ray acceleration at SNR shocks has
certainly improved considerably in the last few years, as also shown
by the impressive quality of the fits to multifrequency observations
of SNRs from 
radio to gamma ray wavelengths. However, it may be useful to briefly
discuss some aspects of the problem that are not so well understood
and that are nevertheless crucial to explain observations and to have
a consistent picture of the origin of cosmic rays. I will limit myself
to mentioning three of such aspects. 

\begin{itemize}
\item{1)} A general trend of the non linear theories of particles
  acceleration at modified shocks is to lead to very large total
  compression 
  factors, incompatible with the values ($\sim 7$) which provide a
  good 
  fit to data. It is usually believed that this reduction may be
  attributed to turbulent heating in the precursor
  \cite{mck,bereTH}. However to date we have no reliable theory of how
  to treat this phenomenon, initially put forward to avoid magnetic
  field amplification to $\delta B/B\gg 1$ and currently used also in
  cases where the amplification is strongly non linear. One can adopt
  recipes that qualitatively work, but do not ensure the correctness
  of the quantitative results.  
\item{2)} The flux of particles that may escape the shock region from
  upstream of a modified shock can be an appreciable fraction of the
  energy flux. We do not know as yet how to carry out a detailed
  calculation of this flux, despite the fact that cosmic rays detected
  at the Earth might be dominated by this component rather than by the
  ones trapped behind the shock. The spectrum of particles that escape
  a SNR at each time during the Sedov phase is expected to be very
  peaked at energies around the maximum momentum achievable at that
  specific time. As proposed by \cite{ptuzira} the observed flux might
  be due to an overlap of these peaked spectra integrated on the SNR
  evolution. 
\item{3)} As discussed above, the crucial element which makes cosmic
  ray acceleration possible in SNR shocks is the magnetic field
  amplification. If this process does not take place, SNRs can hardly
  be the sources of Galactic cosmic rays. The amplification may
  however take place in
  different ways (resonant and non resonant streaming instability,
  firehose instability, turbulent amplification, and possibly
  others). Each amplification process leads to a different saturation
  level and different scaling laws with the age of the SNR. More
  specifically, at different times in the SNR evolution, different
  amplification processes may operate. This makes the calculation of
  the actual spectrum of CRs from SNRs highly non trivial. 
\end{itemize}

It is remarkable that all the work done and presented at the
Conference on shock modification and its phenomenological consequences
was mainly concentrated upon non relativistic shocks and more
specifically SNR shocks. 

A paper was presented \cite{niemiec1051} in which the authors stress
the difficulty of accelerating particles at relativistic shocks (in
the test particle approximation), which reflects into
very steep spectra. This finding adds to the fact, which is becoming
increasingly more recognized, that the dependence on ambient
conditions is very important for the case of relativistic shocks. For
instance, even the simple compression of the perpendicular component
of a turbulent magnetic field crossing a shock surface may induce a
substantial steepening of the spectrum in the relativistic case (see
also \cite{revenu}), while being unimportant for non relativistic
shocks.   

\section{Propagation of Cosmic Rays}
\label{sec:prop}

The propagation of cosmic rays in the Galaxy is dominated by diffusion
(possibly anisotropic) and advection if there is a wind blowing
outwards. Qualitatively it is very easy to illustrate what might be
expected. Let us assume that the sources inject cosmic rays into
the ISM with a rate $Q_p(E)\propto E^{-\gamma}$, where the index $p$
stays for {\it primary}. At the {\it leaky box} level, the equation
which describes the stationary density of cosmic rays in the Galaxy is 
$$Q_p(E)=\frac{n_p(E)}{\tau(E)},$$ 
where $\tau(E)$ is the time needed for escaping the Galaxy. If the
escape is solely due to diffusion and the diffusion coefficient scales
as $D(E)\propto E^\alpha$, then $\tau(E)\propto E^{-\alpha}$. It
immediately follows that $n_p(E)\propto Q_p(E)\tau(E)\propto
E^{-(\alpha+\gamma)}$. During propagation in the ISM secondary nuclei
are produced at a rate $Q_s(E)\approx n_p(E) \sigma n_H v(p)\propto
E^{-(\alpha+\gamma)}$ (assuming $v(p)\sim c$). It follows, again at
the leaky box level of approximation, that the equilibrium density of
secondaries is $n_s(E)\approx Q_s(E) \tau(E)\propto
E^{-(\alpha+2\gamma)}$. Therefore the ratio of the secondary to
primary equilibrium densities is $n_s/n_p \propto E^{-\alpha}\propto
1/D(E)$. This is the case of the $B/C$ ratio discussed in \S
\ref{sec:measure}: measuring the energy dependence of the $B/C$ ratio
we can infer the diffusion coefficient, or more in general the escape
time as a function of energy, which scales as $1/D(E)$ if diffusion is
the only process responsible for escape while it is somewhat more
complex if there are other processes involved (for instance advection
in a wind).   

As discussed in \S \ref{sec:measure} the most recent measurements of
the B/C ratio by CREAM, ATIC and TRACER extended to energies of
$100-1000$ GeV/nucleon, thereby providing us with a unique opportunity
to understand what is the rate of escape of cosmic rays as a function
of energy right below the knee. It is worth recalling that the
data points with small error bars at energies smaller than $\sim 30$
GeV/nucleon (see Fig. \ref{fig:atic_bc}) return a slope of the B/C ratio
$\sim 0.6$. The higher energy data presented by the three
Collaborations have too large error bars so far to understand if such
a slope remains unchanged. On the other hand one should recall that if
indeed cosmic rays escaped from the Galaxy proportionally to
$E^{-0.6}$, and normalizing the escaping time in the 10 GeV/nucleon
region, one would infer a too large anisotropy of cosmic rays around
the knee \cite{hillas}. It is therefore likely that somewhere below
the knee the behaviour of the escape time with energy changes to a
flatter behaviour. On the other hand, if this transition exists then
one should expect the appearance of a feature in the all-particle
spectrum, which does not seem to be there. The problem remains open. 

The importance played by diffusion in the investigation of the origin
of cosmic rays is also clear from the wealth of presentations at the
Conference, concerning different ways to treat diffusion. The standard
way to describe diffusion of cosmic rays in the Galaxy has become the
GALPROP code, which was used also in the work presented
by \cite{moska0738} in order to calculate the elemental abundancies
throughout the periodic table. GALPROP is also used to determine
spectrum and spatial distribution of the secondary emissions,
especially radio and gamma, which provide information on the
distribution of magnetic field and gas respectively. 

Some {\it random-walk}-like approaches to diffusive motion have been
presented in \cite{OG1127,OG1087} and then applied for specific
purposes such as the calculation of the diffuse galactic gamma ray
background or the gamma ray emission from diffuse sources such as the
HESS source in the galactic bulge.

The simple arguments reproduced above, which illustrate the basic
aspects of the leaky box approaches, are a wild oversimplification of
a phenomenon which is in fact very complex. Below I will discuss some
of these complications and how to describe at least some of them. 

In general, diffusion acts on top of a regular motion of charged
particles moving in a large scale magnetic field and is due to a
turbulent component $\delta B$. In the case of the Galaxy the large
scale field has a complex structure, made of spiral arms and a poorly
known halo. Diffusion parallel to the background magnetic field is
faster than diffusion in the direction perpendicular to the ordered
field. The difference between the diffusion coefficients in the two
directions (parallel and perpendicular) decreases when the level of
turbulent field increases. 

The parallel diffusion coefficient can be evaluated in the context of
quasi-linear theory provided $\delta B/B\ll 1$. In the non linear
regime there is no definite theory for the determination of the
diffusion coefficients and numerical simulations become invaluable
tools. In the weakly non linear regime analyical techinques can still
be applied and lead to rather unexpected results: for instance a
similar theoretical approach presented by \cite{OG0046} and consisting
of a weakly non linear approach to parallel diffusion, results in a
scaling of the parallel diffusion coefficient with energy as
$D_\parallel\propto E^{0.6}$ despite the fact that the spectrum of
turbulence is Kolmogorov-like. This might have an important impact on
the interpretation of the observed slope in the B/C ratio, though this
slope would still remain incompatible with the observed anisotropy of
cosmic rays at the knee.

For typical magnetic fields of few $\mu G$, as in the Galaxy,
numerical simulations can be used to derive the propagation properties
of cosmic rays down to energies of $10^{14}-10^{15}$ eV. The results
of one such investigation were presented by \cite{daniel0736} and
revealed several interesting new aspects of the problem. The simulation
consists of propagating a large sample of charged particles and
determine their escape time from {\it toy models} of the magnetic
field of the Galaxy. At the same time, in order to better understand
and interpret the results the authors also calculate the diffusion
coefficients (parallel and perpendicular to the direction of the
regular magnetic field). 

The simulations cover the range of turbulence strength $0.5\leq \delta 
B/B\leq 2$ always assumed to be distributed according with a
Kolmogorov spectrum, $\delta B(k)^2 \propto k^{-5/3}$. The authors
find that the energy dependence of the perpendicular and parallel
diffusion coefficients is different: $D_\perp\propto E^{\alpha}$, with
$\alpha\approx 0.5-0.6$ and $D_\parallel\propto E^{1/3}$ (basically
the same slope obtained from quasi-linear theory despite the non
linearity).  

The results of the calculations of the escape times from the Galaxy
clearly illustrate the diffuculty of the problem. The authors
investigated several toy geometries of the regular magnetic field of
the Galaxy, starting with a purely azimuthal field, from which
particle escape can only take place through diffusion perpendicular to
the field (and therefore to the disc). Even in this simple case, the
escape time is affected not only by (perpendicular) diffusion, but
also by drifts induced by gradients in the regular field. The
gradients may be simply the ones associated with the curvature of
field lines but could also be due to gradients along the radial
direction in the disc, and along the z-axis perpendicular to the
disc. The full set of results are discussed in
\cite{daniel0736,danielJCAP}. 

\begin{figure}
\noindent
\includegraphics [width=.5\textwidth]{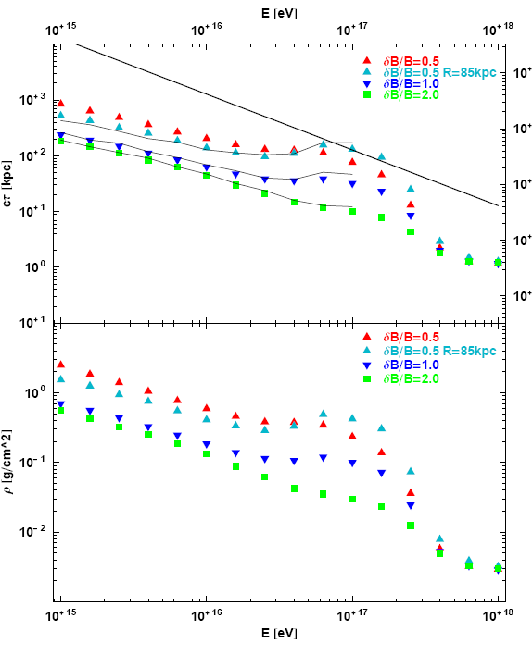}
\caption{Escape times (top panel) and grammage (bottom panel) for
  cosmic rays propagating in a toy model of the Galaxy
  \cite{daniel0736}.}
\label{fig:escape}
\end{figure}

In Fig. \ref{fig:escape} I reproduce the escape times (top panel) and
the grammage traversed by cosmic rays (bottom panel) as presented by
\cite{daniel0736} for the simple case of a purely azimuthal regular
field. A few interesting results are apparent: first, the escape time
obtained with all values of $\delta B/B$ have a slope $\sim 0.6$,
consistent with the fact that the escape time in this field
configuration is $\propto H^2/D_\perp$, where $H$ is the thickness of
the magnetized halo. Second, at $E\sim 10^{17}$ eV the escape time is
appreciably affected by the drift induced by the curvature of the
regular magnetic field lines (solid line). Third, the grammage
inferred from the 
simulation is of order $0.5-3\rm g~cm^{-2}$ at $10^{15}$ eV. If
extrapolated to 10 GeV with a slope 0.6 this would lead to exceedingly
large grammage (or equivalently too long escape times) in the energy
region where this parameter can be inferred from B/C ratio and is
$\sim 20 g~cm^{-2}$. A similar problem, though at higher energies, was 
previously found by \cite{ptuziraprop}. 
Many other cases discussed by the authors confirm the generality of
these few conclusions and show that the problem of propagation has
still many aspects which are poorly understood and can hardly be
included in phenomenological approaches including simple ones such as
the leaky box models or more complex ones such as GALPROP. 
The propagation of ultra-high energy cosmic rays in the intergalactic
space is in a way simpler to describe, though a major source of
uncertainties is due to our ignorance of the magnetic field (strength
and topology) possibly existing between the sources and our Galaxy. 
An appreciable intergalactic magnetic field can affect both the
spectrum and the anisotropy of cosmic rays. While the overall shape of
the spectrum is left almost unchanged, because of the universality
shown by \cite{venya}, some features may appear in the observed
spectrum because of the so-called magnetic horizon \cite{anti,lem}:
at the lowest end of the cosmic ray spectrum, the propagation time
from the nearest source may exceed the age of the universe thereby
suppressing the flux of cosmic rays. For magnetic fields of less than
$\sim 1$ nG, this effect typically appears below $10^{18}$ eV (the
details depend on the topology of the field and the diffusion
properties of the particles). At higher energies the presence of the
magnetic field mainly affects the anisotropy (especially on small
scales) in the arrival directions. In fact at energies $\geq
10^{19.6}$ eV the flux received at Earth is expected to come from
nearby sources, and the arrival directions to point back to the
sources. However there has been a lot of discussion on whether this is
actually the case (see for instance \cite{sigl} and \cite{dolag} for
two rather different results), because of the uncertainties on the
strength of the magnetic field. A paper was presented at the
Conference \cite{kang0418} in which the propagation in a structured
magnetic field was discussed in the assumption that some fraction of
the energy in the hydrodynamic flows leading to large scale structure
formation is converted to magnetic fields. The authors reach a
conclusion which is somewhat in between those of \cite{sigl} and
\cite{dolag}: only some of the UHECRs should point back to their
source. All these results should however be taken with a grain of
salt, the uncertainties in the determination of the magnetic field
being very large. From the observational point of view, if a
correlation was found between arrival directions of UHECRs and a class
of sources, this would be the best proof of the weakness of the
magnetic fields in the intervening intergalactic medium. It would be
harder to conclude one way or the other in case of absence of
correlations. 

The uncertainties in the galactic magnetic field are also rather
disturbing and affect appreciably our ability to obtain realistic
results on the propagation of cosmic rays. In \cite{OG1271} the
authors proposed to use data on the polarization of the CMB photons to
infer information on the large scale structure of the galactic
magnetic field. Such a goal would certainly justify the effort. 

\section{Where do Cosmic Rays become extragalactic?}
\label{sec:trans}

The theoretical arguments illustrated in \S \ref{sec:accel} suggest
that if cosmic rays are accelerated in SNR shocks then the maximum
energy of nuclei of charge $Z$ should be $\sim Z\times 10^{15}$
eV. This conclusion is based on the evidences for magnetic field
amplification at SNR shocks, as inferred from X-ray observations. A
similar conclusion could however be reached based on direct observation
of the proton spectrum by the KASCADE experiment. The proton spectrum
as measured by the Tibet-Array experiment at energies above the knee
is harder than that measured by KASCADE, but the two agree 
on the fact that at $\sim 10^{15}$ eV there is a softening of the
proton spectrum. Moreover, both agree, at least qualitatively, on the 
fact that the chemical composition becomes heavier above the knee in
the all-particle spectrum. 

On the observational side this situation suggests that future accurate 
measurements of the chemical composition at energies above the knee 
will be crucial to solve the problem of the origin of cosmic
rays. From the theoretical side, the most striking conclusion is that
the galactic component of cosmic rays should end at around $\sim \rm
a~few~ 10^{17}$ eV and in this energy region be iron dominated. This is in
clear contradiction with the traditional ankle model of the
transition, which postulates that the transition takes place at $\sim
5\times 10^{18}$ eV as a result of the intersection of a steep
galactic spectrum and a flatter extragalactic (proton dominated)
spectrum. 

Recently two other models have been proposed, both of which locate the
transition region at energies $\sim 10^{17}-10^{18}$ eV, the {\it dip
model} \cite{dip} and the {\it mixed composition} model
\cite{allard}. They are described at length in the rapporteur paper on
session HE and in the Review paper by V.S. Berezinsky, therefore here
I will limit myself to a short description. 

In the dip model \cite{bereall} the extragalactic spectrum is proton
dominated (not more than 15\% of He nuclei are allowed). The effective
injected spectrum at the extragalactic sources is required to be
$E^{-\gamma}$ with $\gamma=2.6-2.7$. The dip appears as a feature due
to pair production losses of protons on the cosmic microwave background
radiation and its location in energy is very well defined and model
independent. The shape of the dip is also very weakly dependent upon
most parameters, with the exception of the chemical composition of the
extragalactic component, as mentioned above. The extragalactic
spectrum extends down to lower energies, and at the position of the
so-called second knee it is predicted to flatten. This is the region
of the transition from a steep galactic to a flatter extragalactic
cosmic ray component. The chemical composition in the transition
region is predicted to suffer a sharp transition from iron dominated
to proton dominated \cite{blasi1}, and the transition is completed at
energy $\sim 10^{18}$ eV. 

In the mixed composition model the extragalactic component corresponds
to a flatter injection spectrum $\sim E^{-2.3}$, with a composition
which is a mixture of different chemicals with abundances that are
roughly comparable with the source compositions inferred for SNRs but
with much freedom in this respect. The propagation of these
components from the sources to Earth leads to the formation of a complex
spectrum that may fit the all particle spectrum fairly well. The
transition from galactic to extragalactic cosmic rays in this model is
more gradual and is completed at energy $\sim 3\times 10^{18}$ eV. The 
chemical composition is expected to vary slowly in the transition
region, shifting from iron dominated (at low energies) to gradually
lighter towards higher energies. 

The main discriminating factor between the two models is the chemical
composition in the transition region. Present measurements do not
allow to make this discrimination as yet. 

In the two models illustrated above, the Galactic cosmic ray component
is obtained by subtraction of the predicted extragalactic flux from
the measured all particle spectrum. In the presentation by \cite{OG1249}
the authors have tried a different approach, namely that of
calculating the galactic component by assuming a source spectrum and
using GALPROP to descrive propagation in the ISM. The main goal was to
show that in general, for both models, the superposition of the
galactic and extragalactic components leads to the appearance of
strong features in the all-particle spectrum. The authors claim that
the features are more evident in the mixed composition scenario. This
result should of course be taken as an interesting suggestion to be
further investigated. We do not have enough information about the
shape of the spectrum of the different chemicals and their maximum
energies in order to run GALPROP and obtain results which can be
interpreted in a unambiguous way. 

\section{Conclusions} 
\label{sec:concl}

The collection of high quality data is the main drive for the field of
cosmic ray research. In the very low energy region (below $\sim 10$
GeV) the accurate measurements of the cosmic ray flux by BESS has,
among other things, allowed to achieve an excellent understanding of
the effects of solar modulation. At energies below the knee in the
all-particle spectrum, several experiments (CREAM, ATIC, Tracer) are
providing us with large statistics of data and correspondingly
excellent quality spectra of nuclei spanning the all periodic
table. The domain of ultra-heavy nuclei is being covered by the TIGER
experiment. 
These direct measurements are gradually approaching the energy
range around the knee thereby providing us with a test of consistency
(or inconsistency) of data collected indirectly by ground experiments
using atmospheric showers induced by the cosmic ray interactions.  
One of the goals of the research in this field in the years to come
should be (and in fact it is already) to match the results of direct
and indirect measurements of cosmic ray fluxes and chemical
composition. Thinking of the possibility of ultra-long duration
balloon flights to reach closer to or across the knee appears as a
worthy effort. 

The achievement of accurate measurements of cosmic ray spectra is also
opening the way to the search for weak signatures of rare phenomena,
such as propagation of antimatter from nearby regions of the universe
and annihilation of dark matter in the Galaxy. The latter has profound
implications on the spectra of positrons and antiprotons especially,
besides creating features in gamma rays and at other wavelengths. The
launch of the PAMELA satellite in 2006 has been a milestone in this
direction, and hopefully next ICRC will have plenty of presentations
of exciting new results. 

The picture that emerges from direct cosmic ray measurements can be
summarized as follows. There is a satisfactory agreement on the
all-particle spectrum 
as measured by different experiments. The spectra of the chemical
species do not show statistically significant differences in the
slopes. The abundances of Ga and Ge do not seem immediately compatible
with acceleration scenarios based on pure first ionization potential
or volatility. Different determinations of the B/C ratio are
consistent with each other up to the highest measured energy ($\sim 1$
TeV), but the error bars are still large enough to leave open the
possibility of a flattening in the slope of the B/C ratio as a
function of energy. A long duration flight, for instance of Tracer,
is likely to settle this issue. We recall that from the theoretical
point of view a naive extrapolation of the observed B/C ratio to the
knee region would lead to exceedingly large anisotropy. 

A few results of measurements carried out with ground experiments, such
as HESS and the Tibet hybrid detector were presented in the OG
session. The HESS collaboration presented the positive detection of
the direct Cherenkov radiation, and used it to infer the spectrum of
iron nuclei (more correctly of nuclei with charge $Z>24$) with energy
up to $\sim 100$ TeV. The measured spectrum appears to be in good
agreement with other results in the same energy region. 

The Tibet Collaboration discussed the results of a measurement of the
proton and helium primary spectra as obtained with the Tibet hybrid
experiment. The discrepancy between these results and those of the
KASCADE experiment are clear, especially for the helium spectrum.  
As stressed above, one way of facing these problems is to extend the
direct measurements with high statistics of events to the knee
region.

The calibration of the ground arrays by an overlap with direct
measurements is a crucial goal to pursue, not only to
understand the origin of the knee but also to describe correctly the
transition from galactic to extragalactic cosmic rays. The two
problems are tightly related to each other. 

There have been numerous presentations at the Conference on the theory
of acceleration of cosmic rays. From the phenomenological point of
view it is clear that several independent pieces support the
possibility that the bulk of galactic cosmic rays are accelerated in
SNRs: 1) X-ray observations have shown that effective magnetic field 
amplification at SNR shocks takes place; 2) multifrequency
observations of SNR RXJ1713 and Vela junior are best fit if the
observed gamma ray emission is of hadronic origin; 3) the spectra of
radio and gamma rays show some evidence, though probably not
conclusive, of curvature (concavity), a phenomenon which is expected
in non linear theories of particle acceleration at shock waves, when
the dynamical reaction of the particles is not negligible, namely when
acceleration is very efficient. 

From the theoretical point of view, the origin of the magnetic field
amplification is being widely investigated. Two mechanisms might be at
work at the same time: streaming instability induced by the
efficiently accelerated cosmic rays and turbulent amplification in a
shocked 
medium with density inhomogeneities. Other instabilities, such as
firehose instability, might also be at work. There have been several
contributed papers on the growth rate of the non resonant streaming
instability in the context of a kinetic approach, and as simulated
with the help of PIC simulations. The latter do not seem to confirm
the large growth rates predicted by quasi-linear theory, but
additional work is needed to make sure that the initial conditions are 
set in a physically meaningful way. 

Efficient magnetic field amplification, when it is due to streaming
instability, also implies efficient particle acceleration, which in
turn induces a concavity in the spectra of both the accelerated
particles and the radiations produced by them. Efficient magnetic
field amplification is also required in order to explain cosmic ray
energies at least in the knee region for protons and correspondingly
higher for nuclei with higher charge. 

All these ingredients are put together in multifrequency
investigations of SNRs, where evidence for magnetic field
amplification comes from X-ray astronomy. Convincing evidence was
presented that such multifrequency spectra, from
radio to gamma rays, can be explained in the framework of particle
acceleration at modified shocks. It was also claimed that the
propagated spectra of different chemical components, when summed
together provide a good fit to the observed all-particle spectrum of
cosmic rays. 

Unfortunately propagation in the Galaxy is all but well understood,
mainly because of our ignorance of both the regular and turbulent
components of the galactic magnetic field: the competition between
parallel and perpendicular diffusion, the presence of a wind, the
topology of the magnetic field in the spiral arms and in between the
arms all affect the escape times and their dependence on energy. A
hint of what might be going on could come from future extensions of
the primary to secondary ratios to higher energies, possibly
approaching the knee region. 

\section{Acknowledgements}

I am grateful to the organizing Committee of the 30th International
Cosmic Ray Conference for their help. I am also grateful to the
numerous scientists in Merida that I had the pleasure to talk to in
order to better prepare this contribution. Finally I wish to express
my gratitude to all my closer collaborators, R. Aloisio, E. Amato,
V. Berezinsky, D. Caprioli, D. De Marco, S. Gabici, G. Morlino,
M. Vietri for continuous stimulating discussions. 

\bibliography{rappox}

\begin{thebibliography}{10}

\bibitem{ahn1173}
H.~S. {Ahn et al.}
\newblock {All-Particle Spectrum Measured by ATIC-1}.
\newblock In {\em ICRC, Merida, Mexico, ICRC.OG.1173}, 2007.

\bibitem{ahn1055}
H.~S. {Ahn et al.}
\newblock {Elemental Spectra from the CREAM-I Flight}.
\newblock In {\em ICRC, Merida, Mexico, ICRC.OG.1055}, 2007.

\bibitem{allard}
{Parizot} E.~{Olinto}~A.~V.~{Kahn}~E. {Allard}, D. and S.~{Goriely}.
\newblock {UHE nuclei propagation and the interpretation of the ankle in the
  cosmic-ray spectrum}.
\newblock {\em Astron. and Astrop.}, 443:29, 2005.

\bibitem{blasi1}
{Berezinsky} V.~{Blasi}~P. {Aloisio}, R. and S.~{Ostapchenko}.
\newblock {Signatures of the transition from galactic to extragalactic cosmic
  rays}.
\newblock {\em Submitted to Phys. Rev. D (arXiv0706.2834)}, 2007.

\bibitem{bereall}
{Berezinsky} V.~{Blasi}~P.~{Gazizov}~A.~{Grigorieva}~S. {Aloisio}, R. and
  B.~{Hnatyk}.
\newblock {A dip in the UHECR spectrum and the transition from galactic to
  extragalactic cosmic rays}.
\newblock {\em Astropart. Phys.}, 27:76, 2007.

\bibitem{anti}
R.~{Aloisio} and V.~S. {Berezinsky}.
\newblock {Anti-GZK Effect in Ultra-High-Energy Cosmic Ray Diffusive
  Propagation}.
\newblock {\em Ap. J.}, 625:249, 2005.

\bibitem{venya}
R.~{Aloisio} and V.~S. {Berezinsky}.
\newblock {Diffusive Propagation of Ultra-High-Energy Cosmic Rays and the
  Propagation Theorem}.
\newblock {\em Ap. J.}, 612:900, 2006.

\bibitem{amato1}
E.~{Amato} and P.~{Blasi}.
\newblock {A general solution to non-linear particle acceleration at
  non-relativistic shock waves}.
\newblock {\em MNRAS Lett.}, 364:76, 2005.

\bibitem{amato2}
E.~{Amato} and P.~{Blasi}.
\newblock {Non-linear particle acceleration at non-relativistic shock waves in
  the presence of self-generated turbulence}.
\newblock {\em MNRAS}, 371:1251, 2006.

\bibitem{bell78}
A.~E. {Bell}.
\newblock {The acceleration of cosmic rays in shock fronts. I}.
\newblock {\em MNRAS}, 182:147, 1978.

\bibitem{bell87}
A.~E. {Bell}.
\newblock {The non-linear self-regulation of cosmic ray acceleration at
  shocks}.
\newblock {\em MNRAS}, 225:615, 1987.

\bibitem{bell2004}
A.~E. {Bell}.
\newblock {Turbulent amplification of magnetic field and diffusive shock
  acceleration of cosmic rays}.
\newblock {\em MNRAS}, 353:550, 2004.

\bibitem{bereTH}
E.~G. {Berezhko} and D.~C. {Ellison}.
\newblock {A Simple Model of Nonlinear Diffusive Shock Acceleration}.
\newblock {\em Ap. J.}, 526:385, 1999.

\bibitem{berez0111}
E.~G. {Berezhko} and H.~J. {V\"{o}lk}.
\newblock {Spectrum of cosmic rays, produced in supernova remnants}.
\newblock In {\em ICRC, Merida, Mexico, ICRC.OG.0111}, 2007.

\bibitem{berez0614}
E.~G. {Berezhko} and H.~J. {V\"{o}lk}.
\newblock {Theory of cosmic ray production in the supernova remnant RX
  J1713.7-3946}.
\newblock In {\em ICRC, Merida, Mexico, ICRC.OG.0614}, 2007.

\bibitem{berez0597}
{Puehlhofer}~G. {Berezhko}, E.~G. and H.~J. {V\"{o}lk}.
\newblock {Cosmic ray acceleration and gamma ray production in the supernova
  remnant RX J0852.0-4622}.
\newblock In {\em ICRC, Merida, Mexico, ICRC.OG.0597}, 2007.

\bibitem{dip}
{Gazizov}~A. {Berezinsky}, V. and S.~{Grigorieva}.
\newblock {Dip in UHECR spectrum as signature of proton interaction with CMB}.
\newblock {\em Phys. Lett. B}, 612:147, 2005.

\bibitem{damiano}
{Amato}~E. {Blasi}, P. and D.~{Caprioli}.
\newblock {The maximum momentum of particles accelerated at cosmic ray modified
  shocks }.
\newblock {\em MNRAS}, 375:1471, 2007.

\bibitem{blasi0342}
P.~{Blasi} and E.~{Amato}.
\newblock {A kinetic approach to non resonant modes and growth rates of
  streaming instability: consequences for shock acceleration}.
\newblock In {\em ICRC, Merida, Mexico, ICRC.OG.0342}, 2007.

\bibitem{blasi0341}
P.~{Blasi} and E.~{Amato}.
\newblock {Theory of non linear particle acceleration at shocks and
  self-generation of the magnetic field}.
\newblock In {\em ICRC, Merida, Mexico, ICRC.OG.0341}, 2007.

\bibitem{boyle1192}
P.~J. {Boyle et al.}
\newblock {Cosmic Ray Energy Spectra of Primary Nuclei from Oxygen to Iron:
  Results from the TRACER 2003 LDB Flight}.
\newblock In {\em ICRC, Merida, Mexico, ICRC.OG.1192}, 2007.

\bibitem{hess0284}
R.~{B\"{u}ler et al. (HESS Coll.)}.
\newblock {Energy spectrum of cosmic iron nuclei measured by HESS}.
\newblock In {\em ICRC, Merida, Mexico, ICRC.OG.0284}, 2007.

\bibitem{coutuetal}
S.~{Coutu et al.}
\newblock {Design and performance in the first flight of the transition
  radiation detector and charge detector of the CREAM balloon instrument}.
\newblock {\em Nucl. Instr. and Meth. A}, 572:485, 2007.

\bibitem{OG1249}
C.~{De Donato} and G.~{Medina Tanco}.
\newblock {The end of the galactic spectrum}.
\newblock In {\em ICRC, Merida, Mexico, ICRC.OG.1249}, 2007.

\bibitem{daniel0736}
{Blasi}~P. {De Marco}, D. and T.~{Stanev}.
\newblock {Numerical Propagation of Cosmic Rays in the Galaxy}.
\newblock In {\em ICRC, Merida, Mexico, ICRC.OG.0736}, 2007.

\bibitem{danielJCAP}
{Blasi}~P. {De Marco}, D. and T.~{Stanev}.
\newblock {Numerical propagation of high energy cosmic rays in the Galaxy: I.
  Technical issues}.
\newblock {\em JCAP}, 6:27, 2007.

\bibitem{OG1127}
{Mastichiadis}~A. {Dimitrakoudis}, S. and A.~{Geranios}.
\newblock {Simulation of Cosmic Ray propagation in the Galactic Centre Ridge in
  Accordance with Observed VHE gamma ray Emission}.
\newblock In {\em ICRC, Merida, Mexico, ICRC.OG.1127}, 2007.

\bibitem{dolag}
{Grasso} D.~{Springel}~V. {Dolag}, K. and I.~{Tkachev}.
\newblock {Constrained simulations of the magnetic field in the local Universe
  and the propagation of ultrahigh energy cosmic rays}.
\newblock {\em JCAP}, 1:9, 2005.

\bibitem{OG0446}
{Van Eldik} C.~{Hinton}~J.~(HESS~Collaboration) {Egberts}, K.
\newblock {Measurement of cosmic ray electrons with HESS}.
\newblock In {\em ICRC, Merida, Mexico, ICRC.OG.0446}, 2007.

\bibitem{heao3}
J.~J. {Engelmann et al. (HEAO Coll.)}.
\newblock {\em Astron. and Astrop.}, 233:96, 1990.

\bibitem{funk}
S.~{Funk}.
\newblock {VHE Gamma-ray supernova remnants}.
\newblock {\em astro-ph/0701471}, 2007.

\bibitem{hams1119}
T.~{Hams et al.}
\newblock {Results from BESS-Polar I 2004 Antarctica Flight}.
\newblock In {\em ICRC, Merida, Mexico, ICRC.OG.1119}, 2007.

\bibitem{hillas}
M.~{Hillas}.
\newblock {Can diffusive shock acceleration in supernova remnants account for
  high-energy galactic cosmic rays?}
\newblock {\em J. of Phys. G}, 31:95, 2006.

\bibitem{OG1087}
C.-Y {Huang} and M.~{Pohl}.
\newblock {Monte Carlo Study of Cosmic-Ray Propagation in the Galaxy and
  Diffuse Gamma-Ray Production}.
\newblock In {\em ICRC, Merida, Mexico, ICRC.OG.1087}, 2007.

\bibitem{OG1271}
{Waelkens} A.~H.~{Farrar}~G.~R. {Jansson}, R. and T.~A. {Ensslin}.
\newblock {Large scale magnetic field of the MilkyWay from WMAP3 data}.
\newblock In {\em ICRC, Merida, Mexico, ICRC.OG.1271}, 2007.

\bibitem{perp}
J.~R. {Jokipii}.
\newblock {Cosmic-Ray Propagation. I. Charged Particles in a Random Magnetic
  Field}.
\newblock {\em Ap. J.}, 146:480, 1966.

\bibitem{joki0078}
J.~R. {Jokipii} and J.~{Giacalone}.
\newblock {Effects of Large-Scale Upstream Turbulence on a Supernova Blast
  Wave}.
\newblock In {\em ICRC, Merida, Mexico, ICRC.OG.0078}, 2007.

\bibitem{kang0418}
{Das} S.~{Ryu}~D. {Kang}, H. and J.~{Cho}.
\newblock {Propagation of UHE protons through a magnetized large scale
  structure}.
\newblock In {\em ICRC, Merida, Mexico, ICRC.OG.0418}, 2007.

\bibitem{lc83a}
P.~O. {Lagage} and C.~J. {Cesarsky}.
\newblock {Cosmic-ray shock acceleration in the presence of self-excited
  waves}.
\newblock {\em Astron. and Astrop.}, 118:223, 1983.

\bibitem{lc83b}
P.~O. {Lagage} and C.~J. {Cesarsky}.
\newblock {The maximum energy of cosmic rays accelerated by supernova shocks}.
\newblock {\em Astron. and Astrop.}, 125:249, 1983.

\bibitem{lem}
M.~{Lemoine}.
\newblock {Extragalactic magnetic fields and the second knee in the cosmic-ray
  spectrum}.
\newblock {\em Phys. Rev. D}, 71:3007, 2005.

\bibitem{revenu}
{Pelletier}~G. {Lemoine}, M. and B.~{Revenu}.
\newblock {On the Efficiency of Fermi Acceleration at Relativistic Shocks}.
\newblock {\em Ap. J. Lett.}, 645:129, 2006.

\bibitem{malkov}
M.~A. {Malkov}.
\newblock {Analytic Solution for Nonlinear Shock Acceleration in the Bohm Limit
  }.
\newblock {\em Ap. J.}, 485:638, 1997.

\bibitem{maldru}
M.~A. {Malkov} and L.~O'C. {Drury}.
\newblock {Non linear theory of diffusive acceleration of particles by shock
  waves}.
\newblock {\em Rep. Prog. in Phys.}, 64:429, 2001.

\bibitem{mck}
J.~F. {McKenzie} and H.~J. {V\"{o}lk}.
\newblock {Non-linear theory of cosmic ray shocks including self-generated
  Alfven waves}.
\newblock {\em Astron. and Astrop.}, 116:191, 1982.

\bibitem{moska0738}
I.~V. {Moskalenko} and A.~W. {Strong}.
\newblock {Isotopic composition of cosmic ray sources}.
\newblock In {\em ICRC, Merida, Mexico, ICRC.OG.0738}, 2007.

\bibitem{niemiec1047}
J.~{Niemiec} and M.~{Pohl}.
\newblock {Magnetic turbulence production by streaming cosmic rays upstream of
  SNR shocks}.
\newblock In {\em ICRC, Merida, Mexico, ICRC.OG.1047}, 2007.

\bibitem{niemiec1051}
{Ostrowski}~M. {Niemiec}, J. and M.~{Pohl}.
\newblock {The inefficiency of the first order Fermi process in UHECR
  production at relativistic shocks}.
\newblock In {\em ICRC, Merida, Mexico, ICRC.OG.1051}, 2007.

\bibitem{panov0018}
A.~D. {Panov et al.}
\newblock {Relative abundances of cosmic ray nuclei B-C-N-O in the energy
  region from 10 GeV/n to 300 GeV/n. Results from ATIC-2 (the science ight of
  ATIC)}.
\newblock In {\em ICRC, Merida, Mexico, ICRC.OG.0018}, 2007.

\bibitem{parketal}
I.~H. {Park et al.}
\newblock {Silicon charge detector for the CREAM experiment}.
\newblock {\em Nucl. Instr. and Meth. A}, 570:286, 2007.

\bibitem{park0593}
N.~H. {Park et al.}
\newblock {Relative abundances of heavy ions measured by the CREAM-II silicon
  charge detector}.
\newblock In {\em ICRC, Merida, Mexico, ICRC.OG.0593}, 2007.

\bibitem{picozza0289}
P.~{Picozza et al.}
\newblock {The Physics of PAMELA Space Mission}.
\newblock In {\em ICRC, Merida, Mexico, ICRC.OG.0289}, 2007.

\bibitem{ptuzira}
V.~S. {Ptuskin} and V.~N. {Zirakashvili}.
\newblock {On the spectrum of high-energy cosmic rays produced by supernova
  remnants in the presence of strong cosmic-ray streaming instability and wave
  dissipation}.
\newblock {\em Astron. and Astrop.}, 429:755, 2005.

\bibitem{rauch0187}
B.~F. {Rauch et al.}
\newblock {Measurement of the Relative Abundances of the Ultra-Heavy Galactic
  Cosmic Rays ($30 \leq Z \leq 40$) with TIGER}.
\newblock In {\em ICRC, Merida, Mexico, ICRC.OG.0187}, 2007.

\bibitem{seo0677}
E.~S. {Seo et al.}
\newblock {Cosmic Ray Energetics And Mass (CREAM) Overview}.
\newblock In {\em ICRC, Merida, Mexico, ICRC.OG.0677}, 2007.

\bibitem{OG0046}
A.~{Shalchi} and R.~{Schlickeiser}.
\newblock {Alternative explanation of the abundance ratio of secondary to
  primary galactic cosmic ray nuclei}.
\newblock In {\em ICRC, Merida, Mexico, ICRC.OG.0046}, 2007.

\bibitem{sigl}
{Miniati}~F. {Sigl}, G. and T.~{Ensslin}.
\newblock {Ultrahigh energy cosmic ray probes of large scale structure and
  magnetic fields}.
\newblock {\em Phys. Rev. D}, 70:3007, 2004.

\bibitem{kieda}
{Kieda}~D.~B. {Swordy}, S.~P. and S.~P. {Wakely}.
\newblock {A high resolution method for measuring cosmic ray composition beyond
  10 TeV}.
\newblock {\em Astropart. Phys.}, 15:287, 2001.

\bibitem{jacco}
J.~{Vink}.
\newblock {X-ray high resolution and imaging spectroscopy of supernova remnants
  }.
\newblock In {\em The X-ray Universe 2005}, page 319, 2006.

\bibitem{berevolk}
{Berezhko}~E.~G. {V\"{o}lk}, H.~J. and L.~T. {Ksenofontov}.
\newblock {Magnetic field amplification in Tycho and other shell-type supernova
  remnants}.
\newblock {\em Astron. and Astrop.}, 433:229, 2005.

\bibitem{Trice}
S.~A. {Wissel et al.}
\newblock {The Status of the Track Imaging Cerenkov Experiment}.
\newblock In {\em ICRC, Merida, Mexico, ICRC.OG.0731}, 2007.

\bibitem{OG0732}
S.~A. {Wissel et al. (Veritas Coll.)}.
\newblock {Studies of Direct Cherenkov Emission with VERITAS}.
\newblock In {\em ICRC, Merida, Mexico, ICRC.OG.0732}, 2007.

\bibitem{yoon0778}
Y.~S. {Yoon et al.}
\newblock {H and He spectra from the 2004/05 CREAM-I flight}.
\newblock In {\em ICRC, Merida, Mexico, ICRC.OG.0778}, 2007.

\bibitem{yoshida0892}
H.~S. {Yoshida et al.}
\newblock {Energy spectrum and arrival directions of high-energy electrons over
  100 GeV observed with PPB-BETS}.
\newblock In {\em ICRC, Merida, Mexico, ICRC.OG.0892}, 2007.

\bibitem{zei0301}
R.~{Zei et al.}
\newblock {Preliminary measurements of Carbon and Oxygen energy spectra from
  the second flight of CREAM}.
\newblock In {\em ICRC, Merida, Mexico, ICRC.OG.0301}, 2007.

\bibitem{ptuziraprop}
{Pochepkin}~D.~N.~{Ptuskin}~V.~S. {Zirakashvili}, V.~N. and S.~I. {Rogovaya}.
\newblock {Propagation of ultra-high-energy cosmic rays in Galactic magnetic
  fields}.
\newblock {\em Astron. Lett.}, 24:139, 1998.

\end{thebibliography}

\bibliographystyle{plain}

\end{document}